\documentclass[10pt]{iopart}

\usepackage{iopams}
\usepackage{graphics}
\usepackage{epsfig}

\newcommand{\degree}{\ensuremath{^\circ}}

\begin{document}

\title[Photolithography operated in a ``Talbot-Fabry-Perot'' regime]{Colloidal pattern replication through contact photolithography 
operated in a ``Talbot-Fabry-Perot'' regime}

\author{Aline Emplit, Jian Xiang Lian, Isabelle Huynen, Alexandru Vlad}
\ead{alexandru.vlad@uclouvain.be}
\address{Institute of Information, Communication Technologies, Electronics and Applied Mathematics, Electrical Engineering, Universit\'{e} catholique de 
Louvain, B-1348 Louvain-la-Neuve, Belgium}
\author{Micha\"{e}l Sarrazin}
\ead{michael.sarrazin@unamur.be}
\address{Research Center in Physics of Matter and Radiation (PMR),
Department of Physics, University of Namur, 61 rue de Bruxelles, B-5000 Namur, Belgium}

\begin{abstract}
We detail on a continuous colloidal pattern replication by 
using contact photolithography. Chrome on quartz masks are fabricated using
colloidal nanosphere lithography and subsequently used as photolithography
stamps. Hexagonal pattern arrangements with different dimensions ($980$, $620$
and $480$ nm, using colloidal particles with respective diameters) have been
studied. When the mask and the imaged resist layer were in intimate contact,
a high fidelity pattern replica was obtained after photolithography exposure
and processing. In turn, the presence of an air-gap in between has been
found to affect the projected image onto the photoresist layer, strongly
dependent on the mask feature size and air-gap height. Pattern replication,
inversion and hybridization was achieved for $980$ nm-period mask; no
hybridization for the $620$ nm; and only pattern replication for the $480$ nm.
These results are interpreted in the framework of a ``Talbot-Fabry-Perot'' effect.
Numerical simulations corroborate with the experimental findings providing
insight into the involved processes highlighting the important parameters
affecting the exposure pattern. The approach allows complex subwavelength patterning 
and is relevant for a 3D layer-by-layer printing.
\end{abstract}


\maketitle

\section{Introduction}

Colloidal particle assisted nanostructuring has become a powerful tool in
nanotechnology \cite{1,2,3,4,5,6,7}. The common approach involves sacrificial use of a
colloidal monolayer and subsequent pattern transfer \textit{via} surface or bulk
structuring. While this approach has provided valuable results and access to
structuring techniques and configuration otherwise difficult to attain, the
original colloidal mask is destroyed and has to be fabricated repeatedly
\cite{8,9,10,11}. Many approaches have been developed for large-scale and easy
assembly of colloidal particles yet the one-time use methodology renders
the approach less attractive. For example, nanoimprint lithography and
photolithography are of interest for industrial applications because they
allow repeated utilization of the mask and continuous processing. Merging
colloidal lithography with one of these techniques is interesting because
the cost for the mask (master) fabrication could be diminished by using
self-assembly techniques whereas the mask could be repeatedly used
replicating the original colloidal pattern.

In this article, we explore this route and find that hexagonal arrangements of circular patterns can be
easily produced by using colloidal-templated Cr-on-quartz
photolithography mask. As such, the colloidal self-assembly is performed
only once, whereas it allows the pattern photo-replication for virtually an
infinite number of times. The mask feature size and arrangement are defined
by the colloidal lithography mask (diameter of the pristine particles
defines the spacing between apertures whereas the size reduction step sets the opening
diameter). The photolithography in hard-contact mode resulted in a series of
patterns, dependent on the mask feature dimensions and especially on the
air-gap distance between the mask and the imaging plane (photoresist layer).

A Talbot-like effect is at the origin of those patterns. The Talbot effect,
also known as the self-imaging effect, consists in periodic repetition of an
image projected through a diffraction grating \cite{18,19,20,21}. When a plane wave is
transmitted through a periodic diffraction grating, repetitive images of the
grating itself are obtained. The Talbot distance is then defined as the
distance from the source where the first regular image is projected, also called
Talbot image. The Talbot imaging has a fractal nature with intermediate
images appearing at smaller distances than the Talbot distance. These
sub-images are characterized by smaller feature size and pattern
hybridization. Moreover, some works consider the Talbot distance in
photolithography. For instance, the Talbot effect can be used to produce
complex 3D periodic patterns \cite{19,21}. Nevertheless, in the present
work, while the typical Talbot distance should be in the micrometer range,
we show that complex patterns occur for small air-gap distance. This is due
to the fact that Cr-on-quartz mask and silicon substrate of the photoresist
layer act as a Fabry-Perot cavity which squeezes the Talbot effect. In the
following section II, the experimental setup is described. Results are
introduced and discussed thanks to numerical simulations in section III.
\newpage

\section{Experimental section}

\begin{figure}[ht]
\centerline{\includegraphics[width=9 cm]{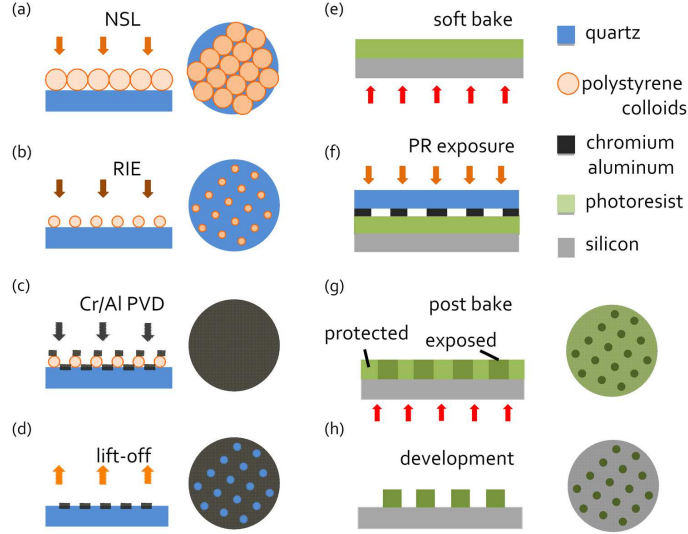}}
\caption{(Color online). Schematic view of the experiments realized in this
work.}
\label{fig1}
\end{figure}

The experimental part of this work is schematized in Fig.1. Polystyrene
colloidal particles with different nominal diameters have been used in this
study as received (Microparticles GmbH; $480$ nm, $620$ nm and $980$ nm). Nanosphere
lithography (NSL, Fig.1a) has been performed using previously documented
techniques on quartz wafers \cite{11,12,13}. Reactive ion etching (RIE, Fig.1b) was
used to reduce the size of the spherical particles. Next, $\approx 80$ nm of metallic Cr was deposited by physical
vapor deposition (PVD) followed by lift-off by adhesive tape (Figs.1c and 1d). To remove any residual organic layer, samples were sonicated in
dichloromethane and exposed to a short O$_2$ plasma ashing. At this stage,
the Cr-on-quartz mask was ready for use.

For the photolithography
experiments, $180$ nm of KMPR-series negative resist was coated on Si chips. In order to attain such thin photoresist films, the KMPR1025 commercially available 
formulation (MicroChem Corp.) was diluted (1/9 volumetric ratio) with cyclohexanone (Sigma Aldrich, b.p. 156\degree C). No
anti-reflective coating layer was used. Prior to the irradiation, the resist
was soft-baked at $100$\degree C for $20$ s (Fig.1e). 
The exposure was done using a K$\&$W Karl S\"{u}ss mask aligner operated in hard-contact mode (Hg-lamp, i-line, cold mirror filtered; exposure wavelength 
centered at 
$365$ nm, $205$ W power) (Fig.1f). The exposure time was varied between $2$ and $3$ s depending on mask type. The resist was then post-baked at $100$\degree C 
for $15$ s and the imaged pattern was revealed in
a standard developing formulation (AZ 726 MIF) for $30$ s at ambient temperature (Figs.1g and 1h). The rather short soft-baking time ($20$ s at $100$\degree 
C) was found insufficient to completely remove the solvent and the coated films were occasionally found to be adherent and easy to indent (see further 
discussions). Nonetheless, the detailed protocol was found to be optimal for attaining high photolithography contrast and resolution. 

Morphology characterization was performed using scanning electron microscopy (SEM, XL 30 FEG) and
optical microscopy. Maps of the square modulus of the electric field are
obtained from a homemade numerical code which uses a Rigorous Coupled Wave
Analysis (RCWA) approach jointly with a scattering matrix method \cite{14,15,16}.

\section{Results and discussions}

\subsection{Description of the procedure}

A schematized view of the experimental approach used in this work is shown in
Fig.1. As can be noticed, the NSL is done only once, during the mask
fabrication step and the fabricated Cr-on-quartz mask is used repeatedly to
replicate the NSL pattern. We have used the fabricated mask for more than 20
times without any sign of degradation. The only required precaution was the
solvent rinsing of the mask after each utilization to remove the photoresist
residues found to affect the quality of the photolithography process. 
We choose to work with NSL for mask fabrication as it allowed us to rapidly fabricate and scan the influence of the feature size of the mask (just by simply 
using colloids with different size) on the photolithography outcome. Furthermore, photolithography masks are usually fabricated using electron-beam lithography 
(especially when sub-micrometer features are targeted), which is a serial writing technique and time consuming. In contrast, colloidal lithography as a bottom-up 
approach can be performed over large areas, in relatively short time and at low costs. Whilst the drawback is the constrained pattern design, some applications 
require just periodic nano-structuring without any special pattern designs like antireflective and super-hydrophobic surfaces or mass production of 
nanostructures for various applications \cite{xx1, xx2, xx3}.

\begin{figure}[ht]
\centerline{\includegraphics[width=9 cm]{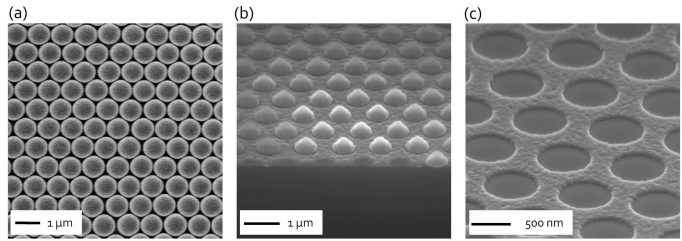}}
\caption{SEM view of the (a) colloidal close packed monolayer, (b) size
reduction using RIE and metal deposition and (c) finished Cr-on-quartz after
sphere lift-off.}
\label{fig2}
\end{figure}

\subsection{Mask fabrication}

Colloidal particles monolayer with hexagonal symmetry is obtained after NSL.
Top-view of an assembled monolayer composed of colloidal particles with a
diameter of $980$ nm is shown in Fig.2a. After the O$_2$ RIE step, the close-packing of the spheres
is altered while the hexagonal arrangement is
preserved (Figs.2b and 2c). After Cr-metal deposition and lift-off, a holey
metal film is obtained. The elevation of the quartz plane with respect to
the photoresist plane is expected to be of the order of $100$ nm (corresponding
to the thickness of the Cr layer). It is worth mentioning here that often, photoresist residues were found on the quartz areas on mask as well as pale 
iridescence was 
observed on the photoresist layer after exposure without being developed (data not shown here). This presumably implies Cr features percolating partially into 
the photoresist layer or, in other words, photoresist and the quartz plane are brought in intimate contact. Hard contact mode used for the lithography step, soft 
nature of the photoresist layer as well as partial bake-out may be at the origin of these experimental observations. However, the thickness of the Cr
layer is only $100$ nm and, as it will be shown further, the critical separation
is much higher than the above mentioned value. Depending on the type of resist (positive or negative), direct or inverted replica respectively, can
be obtained after exposure and revealing. Here, we have used negative tone
photoresist (KMPR series). Consequently, holes of the Cr-on-quartz mask will
normally yield photoresist posts after processing. To avoid thin-film
interference, the thickness of the photoresist film was set to $180$ nm. As
such, the pattern obtained after processing can be approximated to the plane
projected image of the light passing through the holey-metal Cr-on-quartz
mask.

\begin{figure}[ht]
\centerline{\includegraphics[width=8cm]{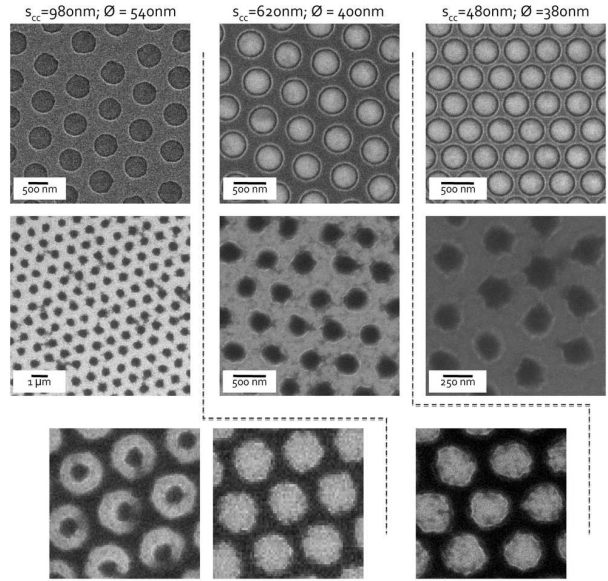}}
\caption{(Top row) SEM views of the fabricated Cr-on-quartz masks using
different diameter colloidal particles. (Middle row) Corresponding SEM
images of the photolithography processing results (Bright areas: photoresist). Typically, the
photoresist dots have been found to replicate with high fidelity the mask
configuration. (Bottom row) Inversed or combined structures have been obtained only for
higher dimensionality masks, i.e. for $S_{cc}$ of $980$ and $620$ nm.}
\label{fig3}
\end{figure}

Examples of fabricated masks with different feature parameters are shown in
Fig.3. Three mask types have been used, fabricated using different colloid
diameters: mask M1, lattice constant $S_{cc}=980$ nm and hole
diameter $\emptyset =540$ nm; mask M2, $S_{cc}=620$ nm and $\emptyset =400$ nm;
mask M3, $S_{cc}=480$ nm and $\emptyset =380$ nm. Careful inspection over the entire area of the mask showed that there are no inverted patterns (i.e. metallic 
dots) as well as the absence
of any impurities at the center of the holes. These observations are
important because the inspection of the photolithography results revealed
the presence of unexpected patterns, especially for M1, not due to the
presence of defects on the mask. Those surprising patterns are related to the Talbot effect, as discussed hereafter.

\subsection{Lithography results}

The photolithography results are shown in Fig.3 (bottom row). Arrays of resist dots have been primarily obtained,
consistent with the processing conditions (openings in a Cr-on-quartz mask
and negative tone photoresist). The dots have circular symmetry and preserve
the hexagonal arrangement for all three types of masks. While the dot lattice spacing is fixed as defined by the imaging mask, their diameter can
be modified by adjusting the exposure dose and post processing parameters
(developer concentration and time). Using M3 mask (Fig.3, bottom row), the diameter of the dots is close to the resolution limit of $\lambda /2\approx 200$ nm. 
Indeed, as shown hereafter, the photoresist layer 
is still located in the radiative near-field zone - i.e. the Fresnel zone - and not in the non-radiative near-field region. As such, simple mask fabrication 
procedure combined with standard photolithography enabled fast and large-area replication of $200$ nm features. 

Another interesting finding was the observation of inverted
and multiplied (along with normal) patterns after photolithography
processing. Examples are shown in the bottom row of Fig.3. Such feature 
was not observed for the mask M3 for which only photoresist posts were obtained. Using mask
M2, pattern inversion was observed: instead of obtaining photoresist posts,
holes in the photoresist films were obtained. Furthermore, three types of
patterns were obtained using the M1 mask: normal - photoresist posts,
inverted - holey photoresist film and a combination of both - i.e., holey
films with posts at the interior of hole (donut-like structures). It should
be noted here that even though hard contact mode photolithography was used,
due to the processing defects and non-planarity of both mask and carrier
sample, conformal contact between both elements was not possible to achieve
over the entire area of the sample. The presence of an air-gap in between
was visually identified by the appearance of the interference patterns
(however, limited to less than $30\%$ of the entire sample area). The
above-mentioned pattern shape anomalies were mainly observed in the
respective areas and seemingly attributed to the air-gap. We did not
performed any extended experiment to precisely control the gap-spacing
between the imaging (Cr-on-quartz mask) and the imaged (photoresist layer)
planes. While this could be implemented easily by deposition of a transparent
dielectric layer \cite{19} of desired thickness either on mask or on top of the
photoresist, we centered our interest in numerically interpreting the
obtained results and finding the critical parameters that affect the
photolithography processing outcome \cite{21}.

\begin{figure}[ht]
\centerline{\includegraphics[width=8 cm]{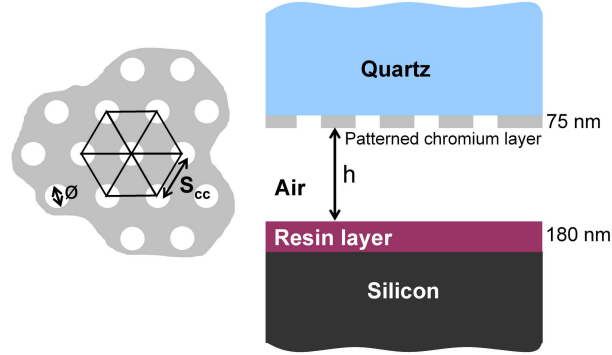}}
\caption{(Color online). Schematics of the configuration employed for the simulations.}
\label{fig4}
\end{figure}

\subsection{Modeling}

Simulations provide the square modulus of the electric field (i.e. the field
intensity) and are obtained from a homemade numerical code which uses a RCWA approach jointly with a scattering
matrix method \cite{14,15,16}. The simulation configuration is depicted in Fig. 4. The
Cr-on-quartz mask is numerically represented as an opaque film bearing
hexagonally arranged holes. The thickness of the Cr layer is set to $75$ nm
while the quartz is being considered of infinite thickness. Similarly, the
silicon supporting substrate is simulated of infinite thickness while the
photoresist is $180$-nm-thick. The air-gap ($h$) is defined as the distance
between the upper surface and lower boundary of the photoresist and
Cr layer, respectively. Assuming that $\lambda =365$ nm, the optical
constant of the materials are as follows: $n_{Cr}=1.40+i3.26$, $%
n_{quartz}=1.57$, $n_{Si}=7.26+i1.27$, $n_r=1.61+i5\cdot 10^{-4}$ for
chromium, quartz, silicon, \cite{17} and resin (KMPR) respectively. In each figure, $h$ is the
separation between the chromium/air and the resin/air interfaces in the
modelized system (see Fig.4). In each figure, the electromagnetic
field intensity scales are given for an incident electric field (in quartz
substrate) equal to $1$ V$\cdot $m$^{-1}$.

\begin{figure}[ht]
\centerline{\includegraphics[width=12 cm]{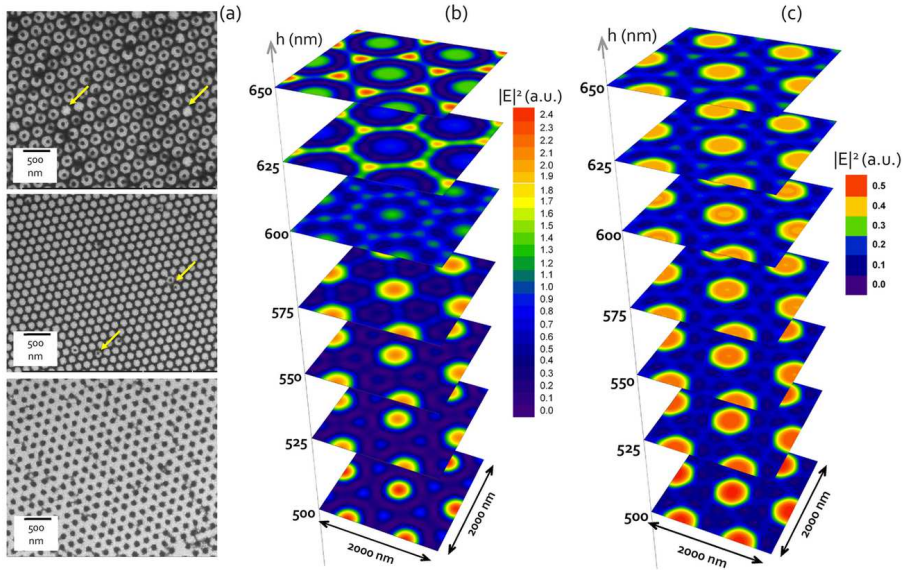}}
\caption{(Color online). (a) SEM images depicting the obtained morphologies using a Cr-on-quartz 
mask with $S_{cc}=980$ nm and $\emptyset =540$ nm. Inversion (middle SEM image) or
hybridization (top SEM image) of the pattern was detected. Occasionally, pattern mixing
(marked with arrows) was observed for higher $h$ structures. (b)
Simulated electromagnetic field intensity at the surface of the photoresist
layer for different $h$ values. 
(c) Simulated profile for the same configuration except that the silicon substrate has been omitted.}
\label{fig5}
\end{figure}

Fig.5 shows the simulation and experimental results for process using
mask M1. For an air-gap of $550$ nm or below, direct imaging is observed,
i.e. arrays of photoresist posts (bottom SEM image). It should be mentioned that no other
patterns were observed in the respective areas. Simulation shows that the
reversed pattern (middle SEM image) is obtained for $h$ around $625$ nm
whereas the hybridized pattern (top SEM image) is recovered for $h$ about $%
650$ nm. All other $h$ values were found to recover the bottom SEM image.
Interestingly, a small variation in $h$ ($\approx 25$ nm) is found to
dramatically affect the projected image requiring precise mask and air-gap
engineering/design if the respective patterns are targeted. Close inspection
of the inverted and hybridized pattern area revealed intermixing of the
respective patterns within. For example, in the inverted pattern (holey
polymer film), occasionally, polymer posts were found in the middle of the
hole and inversely, the absence of the polymer post was observed in the
hybridized pattern (marked with arrows in Fig.5). The very small separation $h$
required to differentiate the two patterns and possible non-uniformities of
the mask features (for example, small variation in the colloid size leading
to different hole size) are the reason for the pattern mixing.

Fig.5c shows the simulation results for conditions similar to those
presented in Fig.5b, except that the silicon substrate and the resin layer
were numerically removed in order to propagate the electromagnetic field in
true Talbot conditions \cite{21}. We clearly see that the electromagnetic field
patterns strongly differ from those obtained in Fig.5b. This shows that
the present experimental results depend on the interferences between the
electromagnetic wave arising from the chromium mask and the reflected wave
on the silicon-resin interface, which acts as a mirror. It should be noted that, in the experiments, no
anti-reflective layer was used on silicon substrate. Eventually, using such
a layer would eliminate the mirror-cavity interferences and lead to direct
pattern replication without any special care with respect to the air-gap
height \cite{19}.

\begin{figure}[ht]
\centerline{\includegraphics[width=8 cm]{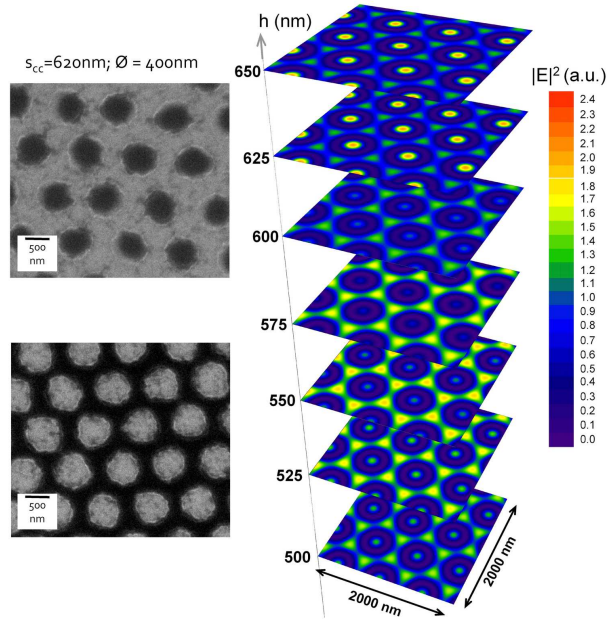}}
\caption{(Color online). (Left) SEM images depicting the obtained morphologies using a Cr-on-quartz 
mask with $S_{cc}=620$ nm and $\emptyset =400$ nm. Pattern
inversion was detected only in this case. (Right) Simulated electromagnetic field
intensity at the surface of the photoresist layer for different $h$ values.}
\label{fig6}
\end{figure}

Having confirmed the applicability of the simulation protocol to the
observed experimental results, we have analyzed the results obtained using
masks M2 and M3. Fig.6 shows the experimental results obtained using mask
M2 and the corresponding simulated electromagnetic field intensity at the
surface of the photoresist layer. The inverted pattern, i.e. holey
photoresist film (bottom SEM image), is roughly recovered for $h$ between $500$
 and $600$ nm. The top SEM image (the normal pattern) is obtained for the
other $h$ values (higher than $600$ nm and lower than $500$ nm). The pattern
mixing was not observed here, neither experimentally nor in the simulations.
The results for mask M3 are shown in Fig.7. In this configuration, only
direct patterning was observed. Polymer posts were obtained independently of
the $h$ value.

\begin{figure}[ht]
\centerline{\includegraphics[width=8 cm]{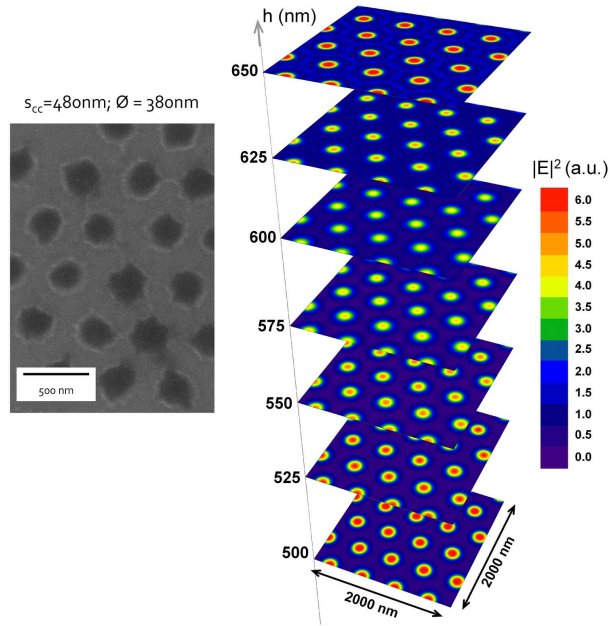}}
\caption{(Color online). (Left) SEM images depicting the obtained morphologies using a Cr-on-quartz 
mask with $S_{cc}=480$ nm and $\emptyset =380$ nm. Neither pattern
mixing or inversion was observed for higher $h$ values. (Right)
Simulated electromagnetic field intensity at the surface of the photoresist
layer for different $h$ values.}
\label{fig7}
\end{figure}

\begin{table}[ht]
\begin{center}
\begin{tabular}{llll}
$M1$ & $Z_T^{\prime } (nm)$ & $h_{FP} (nm)$ & $h (nm)$ \\ \hline\hline
 & 751 & 730 & 750 \\ 
 & 536  & 548 & 550 \\ 
 & 417  & - & - \\ 
 & 341 & 365 & 350 \\
 & 289 & - & - \\
 & 250 & - & - \\
 & 221 & - & - \\
 & 198 & - & - \\
 & 179 & 183 & 200 \\
$M2$ & $Z_T^{\prime} (nm)$ & $h_{FP} (nm)$ & $h (nm)$ \\ \hline\hline
 & 456 & 548 & Undefined \\ 
 & 274 & 365 & 350 \\
 & 196 & 183 & 200 \\
$M3$ & $Z_T^{\prime} (nm)$ & $h_{FP} (nm)$ & $h (nm)$ \\ \hline\hline
 & 700 & 730 & Undefined\\
 & 233 & 183 & Undefined\\ 
\end{tabular}
\end{center}
\caption{Comparison between theoretical squeezed Talbot distances $Z_T^{\prime}$, Fabry-Perot wavelengths $h_{FP}$ and numerical Talbot distances $h$ in the 
range $h=200$ - $800$ nm.}
\label{photocurrent}
\end{table}

\subsection{Discussions}

As previously mentioned, the Talbot effect consists in periodic repetition of
an image of a diffraction grating \cite{18,19,20}. The Talbot distance is the
distance between the source and the first regular image. The Talbot imaging
presents intermediate images appearing at smaller distances than the Talbot
distance. These sub-images are characterized by smaller feature size and
pattern hybridization. The Talbot distance ($Z_T$) can be approximated
using the following equation provided that light propagates in air \cite{20}: 

\begin{equation}
Z_T=\frac \lambda {1-\sqrt{1-(\lambda /a)^2}}\label{1}
\end{equation}

where $\lambda $ is the incident wavelength, $a$ is the grating lateral
period ($a=S_{cc}\sqrt{3}/2$, with $S_{cc}$ the lattice period of the mask).
For the experimental set-up and the three different masks used in this study,
values $Z_T$ are as follows: for mask M1, $Z_T=3.76$ $\mu$m; for mask M2, $%
Z_T=1.37$ $\mu$m; and for M3, $Z_T=0.7$ $\mu$m. These large values lead to
almost constant patterns when the distance $Z$ between mask and photoresist
slowly varies, as shown in Fig.6. Nevertheless, in the present case, the
silicon substrate dramatically modifies the lateral pattern behavior. Indeed,
the Cr-on-quartz mask and the silicon-photoresist support act as a Fabry-Perot 
cavity allowing for fast lateral pattern variation when $Z$ varies. Since the
silicon acts as a mirror, we can fairly suppose that the new Talbot distance (squeezed Talbot distance)
$Z_T^{\prime }$ will be given by $Z_T^{\prime }=Z_T/(2n+1)$ where $n$ is an
integer. Indeed, with $h=Z_T^{\prime }$, which means that the Talbot distance 
$Z_T$ is retrieved with $Z_T=(2n+1)h$, i.e. when light has made $2n+1$ trips
between the source (the mask) and the mirror (the silicon substrate). An odd
number of trips is required to get the final image of the source on the mirror. 
For each mask, the values of $Z_T^{\prime}$ are shown in Table 1.
In addition, the Fabry-Perot implies that the distance $h_{FP}$ for which
the field can resonantly propagate between both plates is given by $%
h_{FP}=m\lambda /2$ ($m$ is an integer different from zero). Those values 
can be compared with the theoretical squeezed Talbot distance $Z_T^{\prime}$ as shown in Table 1.
Then, for each mask, we can compare the numerically computed squeezed Talbot distance with 
the relevant values of $Z_T^{\prime}$ (see Table 1).

For masks M1 and M2, this approach matches quite well. For mask M3, we are closer to the true Talbot conditions and we lose the capability to get complex 
subwavelength patterns. Typical air-gap widths are between $350$ et $700$ nm, i.e. $Z_T$ and $Z_T/2$, and the mask pattern is retrieved all along those 
distances. 
Pattern deformation during photolithography induced by the presence of mask-substrate gaps is a known phenomenon and several studies have analyzed this effect 
and solutions to compensate the distortions have been proposed \cite{xxx}. The major difference with respect to data presented in this work is that the air-gaps 
utilized in previous reports as well as characteristic feature dimensions were far much higher than the exposure wavelength. The documented distortions were 
attributed to the divergence of the exposure source (in the combined optical system made of light source/mask/photoresist plane). Moreover, in our case, 
the characteristic dimensions are of the order of the wavelength and the source divergence is expected to have a minimal influence on the results. 
At last, note that, even though the real part of the Cr permittivity is negative here, surface plasmons cannot be excited. Indeed, since the incident 
light is at normal incidence, any surface plasmon excitation is due to diffraction orders which become tangent to the metallic surface. In the present case, 
for the hexagonal lattice, no orders can provide a resonant excitation for wavelengths in the very close ``vicinity'' of $365$ nm whatever the lattice 
parameter. As a consequence, surface plasmons play no role in the present study.

\section{Conclusions}
We have introduced a continuous colloidal pattern replication 
using contact photolithography thanks to a Talbot-like effect.
Cr-on-quartz masks have been fabricated using nanosphere lithography and used as photolithography
stamps. Dimensions of the pattern are experimentally controllable (initial size of the colloids, the etching procedure, etc.). 
Hexagonal pattern arrangements with different dimensions have been considered. Depending on the thickness of the air-gap between the mask and the photoresist 
layer, various changes occur in the pattern replica. The present approach then allows to obtain complex bidimensional patterns at a subwavelength scale. For 
instance, not only dots can be obtained (high fidelity replica of the mask), but also subwavelength anti-dots or rings. As a consequence, this patterning 
approach benefits from the ``infinite'' use of the mask while allowing multi-pattern with a single mask.
Numerical simulations and a simple model allow to predict the obtained patterns against the air-gap thickness.
Further improvements could be considered with a piezoelectric actuator allowing for an accurate control of the air-gap and, then, of the obtained patterns. 
This could also allow 3D layer-by-layer printing.

A.V. acknowledges F.R.S.-FNRS for financial support. This research used resources of the ``Plateforme Technologique de Calcul Intensif (PTCI)''
(http://www.ptci.unamur.be) located at the University of Namur, Belgium, which is supported by the F.R.S.-FNRS under the convention No. 2.4520.11. The PTCI is 
member of the ``Consortium des \'{E}quipements de Calcul Intensif (C\'{E}CI)'' (http://www.ceci-hpc.be).


\section*{References}
\begin{thebibliography}{10}

\bibitem{1}  T. Kraus, D. Brodoceanu, N. Pazos-Perez, A. Fery, \textit{Colloidal
Surface Assemblies: Nanotechnology Meets Bioinspiration}, Advanced Functional
Materials \textbf{23} (2013) 4529.

\bibitem{2}  C.G. Sch\"{a}fer, M. Gallei, J.T. Zahn, J. Engelhardt, G.P.
Hellmann, M. Rehahn, \textit{Reversible Light-, Thermo-, and Mechano-Responsive
Elastomeric Polymer Opal Films}, Chemistry of Materials \textbf{25} (2013) 2309-2318.

\bibitem{3}  J. Zhang, Y. Li, X. Zhang, B. Yang, \textit{Colloidal Self-Assembly
Meets Nanofabrication: From Two-Dimensional Colloidal Crystals to
Nanostructure Arrays}, Advanced Materials \textbf{22} (2010) 4249-4269.

\bibitem{4}  H. Fredriksson, Y. Alaverdyan, A. Dmitriev, C. Langhammer, D.S.
Sutherland, M. Z\"{a}ch, et al., \textit{Hole-Mask Colloidal Lithography}, Advanced
Materials \textbf{19} (2007) 4297-4302.

\bibitem{5}  H. Zhang, P.V. Braun, \textit{Three-Dimensional Metal Scaffold
Supported Bicontinuous Silicon Battery Anodes}, Nano Letters. \textbf{12} (2012)
2778¤2783.

\bibitem{6}  S. Kim, S. Lee, S. Yang, G.-R. Yi, \textit{Self-assembled colloidal
structures for photonics}, NPG Asia Materials \textbf{3} (2011) 25-33.

\bibitem{7}  J. Zhang, B. Yang, \textit{Patterning Colloidal Crystals and
Nanostructure Arrays by Soft Lithography}, Advanced Functional Materials \textbf{20}
(2010) 3411-3424.

\bibitem{8}  N. Vogel, L. De Viguerie, U. Jonas, C.K. Weiss, K. Landfester,
\textit{Wafer-Scale Fabrication of Ordered Binary Colloidal Monolayers with
Adjustable Stoichiometries}, Advanced Functional Materials \textbf{21} (2011) 3064-3073.

\bibitem{9}  S.P. Bhawalkar, J. Qian, M.C. Heiber, L. Jia, \textit{Development of a
Colloidal Lithography Method for Patterning Nonplanar Surfaces}, Langmuir \textbf{26}
(2010) 16662-16666.

\bibitem{10}  H.-P. Wang, K.-Y. Lai, Y.-R. Lin, C.-A. Lin, J.-H. He,
\textit{Periodic Si Nanopillar Arrays Fabricated by Colloidal Lithography and
Catalytic Etching for Broadband and Omnidirectional Elimination of Fresnel
Reflection}, Langmuir \textbf{26} (2010) 12855-12858.

\bibitem{11}  A. Vlad, A. Fr\"{o}lich, T. Zebrowski, C.A. Dutu, K. Busch, S.
Melinte, et al., \textit{Direct Transcription of Two-Dimensional Colloidal Crystal
Arrays into Three-Dimensional Photonic Crystals}, Advanced Functional
Materials \textbf{23} (2013) 1164-1171.

\bibitem{18}  J. Wen, Y. Zhang, M. Xiao, \textit{The Talbot effect: recent advances
in classical optics, nonlinear optics, and quantum optics}, Adv. Opt. Photon.
\textbf{5} (2013) 83.

\bibitem{19}  C.H. Chang, L. Tian, W.R. Hesse, H. Gao, H.J. Choi, J.G. Kim,
et al., \textit{From Two-Dimensional Colloidal Self-Assembly to Three-Dimensional
Nanolithography}, Nano Letters \textbf{11} (2011) 2533-2537.

\bibitem{20}  J.T. Winthrop, C.R. Worthington, \textit{Theory of Fresnel Images I
Plane Periodic Objects in Monochromatic Light}, J. Opt. Soc. Am. \textbf{55} (1965)
373.

\bibitem{21} T. Sato, \textit{High-Order Approximation of the Talbot Distance for Lithography}, Appl. Phys. Express \textbf{5} (2012) 092501.

\bibitem{12}  A. Vlad, I. Huynen, S. Melinte, \textit{Wavelength-scale lens
microscopy via thermal reshaping of colloidal particles}, Nanotechnology \textbf{23}
(2012) 285708.

\bibitem{13}  N. Reckinger, A. Vlad, S. Melinte, J.-F. Colomer, M. Sarrazin,
\textit{Graphene-coated holey metal films: tunable molecular sensing by surface
plasmon resonance}, Applied Physics Letters \textbf{102} (2013) 211108.

\bibitem{14}  M. Sarrazin, J.-P. Vigneron, J.-M. Vigoureux, \textit{Role of Wood
anomalies in optical properties of thin metallic films with a bidimensional
array of subwavelength holes}, Phys. Rev. B \textbf{67} (2003) 085415.

\bibitem{15}  J.-P. Vigneron, V. Lousse, \textit{Variation of a photonic crystal
color with the Miller indices of the exposed surface}, Proc. of SPIE. \textbf{6128}
(2006) 61281G.

\bibitem{16}  J.-P. Vigneron, F. Forati, D. Andr\'{e}, A. Castiaux, I.
Derycke, A. Dereux, \textit{Theory of electromagnetic energy transfer in
three-dimensional structures}, Ultramicroscopy \textbf{61} (1995) 21-27.

\bibitem{xx1} J.-J. Kim, Y. Lee, H. Kim, K.-J. Choi, H.-S. Kweon, S. Park, K.-H. Jeong, \textit{Biologically inspired LED lens from cuticular nanostructures of 
firefly lantern}, Proc. Natl. Acad. Sci. \textbf{109} (2012) 18674.

\bibitem{xx2} A. Vlad, A.L.M. Reddy, A. Ajayan, N. Singh, J.-F. Gohy, S. Melinte, P.M. Ajayan, \textit{Roll up nanowire battery from silicon chips}, Proc. Natl. 
Acad. Sci. \textbf{109} (2012) 15168.

\bibitem{xx3} K.-C. Park, H.J. Choi, C.-H. Chang, R.E. Cohen, G.H. McKinley, G. Barbastathis, \textit{Nanotextured silica surfaces with robust 
superhydrophobicity and omnidirectional broadband supertransmissivity}, ACS Nano \textbf{6} (2012) 3789.

\bibitem{17}  E.D. Palik, \textit{Handbook of Optical Constants of Solids II}, 1997
Elsevier Inc. (1997). 

\bibitem{xxx} W. J. Venstra, J.W. Spronck, P. M. Sarro, J. van Eijk, \textit{Photolithography on bulk micromachined substrates}, J. Micromech. Microeng. 
\textbf{19} (2009) 055005. 

\end{thebibliography}
\end{document}